# Coverage Probability and Ergodic Capacity of Intelligent Reflecting Surface-Enhanced Communication Systems

Trinh Van Chien, *Member, IEEE*, Lam Thanh Tu, Symeon Chatzinotas, *Senior Member, IEEE*, and Björn Ottersten, *Fellow, IEEE*

*Abstract*—This paper studies the performance of a single-input single-output (SISO) system enhanced by the assistance of an intelligent reflecting surface (IRS), which is equipped with a finite number of elements under Rayleigh fading channels. From the instantaneous channel capacity, we compute a closed-form expression of the coverage probability as a function of statistical channel information only. A scaling law of the coverage probability and the number of phase shifts is further obtained. The ergodic capacity is derived, then a simple upper bound to simplify matters of utilizing the symbolic functions and can be applied for a long period of time. Numerical results manifest the tightness and effectiveness of our closed-form expressions compared with Monte-Carlo simulations.

## I. Introduction

Serious concerns on the spectral and energy efficiency have been raised in wireless communications due to high data rate demand from more than 12 billion wirelessly connected devices by 2022 [1]. Among potential solutions, IRS has recently emerged as an energy-efficient hardware technology to enhance the performance of beyond 5G wireless systems [2], [3]. An IRS is equipped with passive reflecting elements to support reliable communication by controlling how the electromagnetic waves arrive and reflect on its surface. Steering each element independently to get the desired phase shifts, an IRS enables producing coherent signal combining with a constructive gain at the rate of fast fading. Despite a long history in the electromagnetic literature of the meta-surface design [4], system performance analyses in wireless communications are largely unexplored.

An IRS giving unprecedented supports to improve the channel capacity was reported in [5] without using an amplifier, leading to low power consumption compared to a relay. Following, jointly optimizing passive phase shifts in conjunction with other traditionally controlled wireless resources was considered to intensify different utility metrics as energy efficiency [6], max-min fairness [7], and to provide better physical layer security [2] together with references therein.

While those previous works are utterly based on small-scale fading, only a few related works pay attention to system analyses utilizing large-scale fading coefficients, but under the asymptotic regime, such as [8] for a large IRS or [9] for multiple IRSs. In [10], the authors computed the closed-form expression of the coverage property for a large number of the optimal phase shifts and no direct link between the source and destination. However, the radio communication characteristics may behave differently when there is a small finite number of phase shifts.

In this paper, unlike the above-mentioned works, we fill the gap by analyzing the system performance of SISO communication with either the optimal or arbitrary phase shifts at IRS. For such purpose, we compute the coverage probability in a closed-form expression. It matches well to a Gamma distribution whose shape and scale parameters depend on large-scale fading coefficients and the number of phase shifts. From the statistical information of the obtained coverage probability, we also compute the ergodic capacity in a closed-form, which is independent of small-scale fading, and is expressed solidly by a simple fraction between the MeijerG and Gamma functions, instead of infinitive summations as in previous works. For transparent insight and confronting with the literature, an upper bound of the ergodic capacity is then formulated.

*Notations*: The capital and lower bold letters are used for matrices and vectors, respectively. The superscripts $(\cdot)^*$ and $(\cdot)^H$ denote the conjugate and Hermitian transpose, respectively. An identity matrix of size $N \times N$ denotes as $\mathbf{I}_N$; $\|\cdot\|$ is Euclidean norm and $|.|$ is absolute operator. The probability, expectation, and variance of a random variable is $\Pr(\cdot), \mathbb{E}\{\cdot\}$, and $\mathsf{Var}\{\cdot\}$, respectively. A diagonal matrix with $x_1, \ldots, x_N$ on the diagonal denotes by $\mathrm{diag}(x_1, \ldots, x_N)$. Finally, $\mathcal{CN}(\cdot, \cdot)$ is a circularly symmetric complex Gaussian distribution.

## II. System Model and Channel Capacity

We consider a system comprising one single-antenna source communicating with one single-antenna destination under the supports of an IRS equipped with $N$ phase-shift elements and located in the coverage area. We exploit a block-fading channel model where the channels are quasi-static in each coherence interval. In particular, the uncorrelated Rayleigh channel model is used, for which the channel between source and destination is $h_\mathrm{sd} \sim \mathcal{CN}(0, \beta_\mathrm{sd})$; between source and IRS is $\mathbf{h}_\mathrm{sr} \sim \mathcal{CN}(\mathbf{0}, \beta_\mathrm{sr}\mathbf{I}_N)$; between IRS and destination is $\mathbf{h}_\mathrm{rd} \sim \mathcal{CN}(\mathbf{0}, \beta_\mathrm{rd}\mathbf{I}_N)$, where $\beta_\mathrm{sd}, \beta_\mathrm{sr}$, and $\beta_\mathrm{rd}$ are large-scale fading coefficients. The phase-shift matrix is $\boldsymbol{\Theta} = \mathrm{diag}\left(\exp(j\theta_1), \ldots, \exp(j\theta_N)\right)$, where $\theta_n \in [-\pi, \pi], \forall n$, are the phase shifts induced by the IRS. If source is transmitting

T. V. Chien, S. Chatzinotas, and B. Ottersten are with the University of Luxembourg (SnT), Luxembourg (e-mail: vanchien.trinh@uni.lu, symeon.chatzinotas@uni.lu, and bjorn.ottersten@uni.lu).
L. T. Tu is with the institute XLIM, University of Poitiers, Poitiers, France (email: lam.thanh.tu@univ-poitiers.fr).
 

an information symbol $s$ with $\mathbb{E}\{|s|^2\} = 1$ and assuming first-order reflection from IRS only, the received complex baseband signal at destination is

$$r = \sqrt{\rho}\mathbf{h}_{\text{sr}}^H \mathbf{\Theta} \mathbf{h}_{\text{rd}} s + \sqrt{\rho} h_{\text{sd}} s + \tilde{n}, \quad (1)$$

where $\rho$ is the transmit power allocated by source to the information symbol and $\tilde{n} \sim \mathcal{CN}(0, \sigma^2)$ is additive noise. By exploiting the similar methodology as in [5, Sec. III-B], we obtain the channel capacity as in Lemma 1.

**Lemma 1.** *By assuming coherent combination, for arbitrary phase shifts, the instantaneous channel capacity [b/s/Hz] is formulated as*

$$R_a = \log_2\left(1 + \frac{\rho}{\sigma^2}\left|h_{\text{sd}} + \mathbf{h}_{\text{sr}}^H \mathbf{\Theta} \mathbf{h}_{\text{rd}}\right|^2\right), \quad (2)$$

*and for the optimal phase shifts, where $\theta_n = \arg(h_{\text{sd}}) - \arg([\mathbf{h}_{\text{sr}}^*]_n) - \arg([\mathbf{h}_{\text{rd}}]_n), \forall n$, the instantaneous channel capacity [b/s/Hz] is formulated as*

$$R_o = \log_2\left(1 + \frac{\rho}{\sigma^2}\left(|h_{\text{sd}}| + \sum_{n=1}^{N}|[\mathbf{h}_{\text{sr}}]_n||[\mathbf{h}_{\text{rd}}]_n|\right)^2\right), \quad (3)$$

*where $[\mathbf{h}_{\text{sr}}]_n$ and $[\mathbf{h}_{\text{rd}}]_n$ are the $n$-th element of $\mathbf{h}_{\text{sr}}$ and $\mathbf{h}_{\text{rd}}$.*

The channel capacities in Lemma 1 are the function of instantaneous channels varying upon coherence intervals. We hereafter use those results to obtain the coverage probability and ergodic capacity expressed by statistical information only.

## III. Coverage Probability and Ergodic Capacity

This section first derives the coverage probability and ergodic capacity in closed-form expressions. An upper bound of the latter is further obtained by using Jensen's inequality and computing the moments of non-Gaussian random variables.[1]

### A. Coverage Probability Analysis

From the channel capacities presented in (2) and (3), we now consider the coverage probability of the network, which is defined either for the arbitrary or optimal phase shifts as $P_a = 1 - \Pr(R_a < \xi)$, and $P_o = 1 - \Pr(R_o < \xi)$, where $\xi$ [b/s/Hz] is the required threshold information rate. By setting $z = \sigma^2(2^\xi - 1)/\rho$, we recast each coverage probability to an equivalent signal-to-noise ratio (SNR) requirement:

$$P_a = 1 - \Pr\left(\left|h_{\text{sd}} + \mathbf{h}_{\text{sr}}^H \mathbf{\Theta} \mathbf{h}_{\text{rd}}\right|^2 < z\right), \quad (4)$$

$$P_o = 1 - \Pr\left(\left(|h_{\text{sd}}| + \sum_{n=1}^{N}|[\mathbf{h}_{\text{sr}}]_n||[\mathbf{h}_{\text{rd}}]_n|\right)^2 < z\right). \quad (5)$$

In this paper, the moment-matching method is used to manipulate the coverage probabilities in (4) and (5).

**Theorem 1.** *For arbitrary phase shifts, the coverage probability in (4) is computed as*

$$P_a = \Gamma(k_a, z/w_a)/\Gamma(k_a), \quad (6)$$

[1]This paper treats the phase shifts as deterministic parameters when computing the closed-form expressions, thus our framework can be extended to practical IRS constraints such as IRS with discrete phase shifts and/or coupled reflection amplitude and phase shifts.

*where $\Gamma(m, n) = \int_n^\infty t^{m-1}\exp(-t)dt$ is the upper incomplete Gamma function, and $\Gamma(x) = \int_0^\infty t^{x-1}\exp(-t)dt$ is the Gamma function. The shape parameter $k_a$ and scale parameter $w_a$ are*

$$k_a = \frac{(\beta_{\text{sd}} + N\beta_{\text{sr}}\beta_{\text{rd}})^2}{(\beta_{\text{sd}} + N\beta_{\text{sr}}\beta_{\text{rd}})^2 + 2N\beta_{\text{sr}}^2\beta_{\text{rd}}^2},$$

$$w_a = \beta_{\text{sd}} + N\beta_{\text{sr}}\beta_{\text{rd}} + \frac{2N\beta_{\text{sr}}^2\beta_{\text{rd}}^2}{\beta_{\text{sd}} + N\beta_{\text{sr}}\beta_{\text{rd}}}. \quad (7)$$

*For the optimal phase shifts, the coverage probability in (5) is computed as*

$$P_o = \Gamma(k_o, \sqrt{z}/w_o)/\Gamma(k_o), \quad (8)$$

*where the shape parameter $k_o$ and scale parameter $w_o$ are*

$$k_o = \frac{\left(\sqrt{\beta_{\text{sd}}\pi} + 2Nkw\right)^2}{4\beta_{\text{sd}} + 4Nkw^2 - \beta_{\text{sd}}\pi}, w_o = \frac{4\beta_{\text{sd}} + 4Nkw^2 - \beta_{\text{sd}}\pi}{2\left(\sqrt{\beta_{\text{sd}}\pi} + 2Nkw\right)}, \quad (9)$$

*with $k = \pi^2/(16 - \pi^2)$ and $w = (4 - \pi^2/4)\sqrt{\beta_{\text{sr}}\beta_{\text{rd}}}/\pi$.*

*Proof.* The proof is available in Appendix A and B for the case of arbitrary and the optimal phase shifts, respectively. □

Theorem 1 gives the closed-form expressions of the coverage probabilities, which are only the function of large-scale fading coefficients. For a different selection of the phase shifts, it shows that the shape and scale parameters converge to the different points at an asymptotic regime. As $N \to \infty$, we mathematically obtain

$$k_a \to 1, w_a \to N\beta_{\text{sr}}\beta_{\text{rd}}, \quad (10)$$

$$k_o \to 4N^2k^2w^2, w_o \to 1. \quad (11)$$

Considering a large IRS with many phase shifts, the coverage probabilities can be approximated as in Corollary 1.

**Corollary 1.** *As $N \to \infty$, the closed-form expression of the coverage probability for the arbitrary and optimal phase shifts respectively holds that*

$$P_a \to 1 - \frac{z}{N\beta_{\text{sr}}\beta_{\text{rd}}}, P_o \to 1 - \frac{\sqrt{z}}{N^2k^2w^2}. \quad (12)$$

*Proof.* For arbitrary phase shifts, the closed-form expression of the coverage probability in (6) is recast as

$$P_a = 1 - \frac{1}{\Gamma(k_a)}\gamma(k_a, z/w_a) = 1 - \frac{(z/w_a)^{k_a}}{k_a\Gamma(k_a)}\sum_{t=0}^{\infty}\frac{(-z/w_a)^t}{(k_a + t)t!}, \quad (13)$$

where $\gamma(m, n) = \int_0^n t^{m-1}\exp(-t)dt$ is the lower incomplete gamma function. $N \to \infty$ leads to $1/w_a \to 0$ and therefore, the summation only $t = 0$ is remained while the high orders, i.e. $t \geq 1, \forall t$, can be neglected. Mathematically, we obtain the following approximation

$$P_a \to 1 - \frac{(z/w_a)^{k_a}}{k_a^2\Gamma(k_a)}. \quad (14)$$

Utilizing the asymptotic property (10) into (14), we obtain $P_a$ as in the corollary.

For the optimal phase shifts, the closed-form expression of the coverage probability is first reformulated as

$$P_o = 1 - \frac{1}{\Gamma(k_o)}\gamma(k_o, \sqrt{z}/w_o), \quad (15)$$

then the following series of the inequalities are constructed as

$$\gamma\left(k_o, \sqrt{z}/w_o\right) < \int_0^{\sqrt{z}/w_o} \left(\sqrt{z}/w_o\right)^{k_o-1} \exp\left(-\sqrt{z}/w_o\right) dt$$
$$= \frac{\sqrt{z}/w_o}{k_o - 1 - \sqrt{z}/w_o} \int_{\sqrt{z}/w_o}^{k_o-1} \left(\sqrt{z}/w_o\right)^{k_o-1} \exp\left(-\sqrt{z}/w_o\right) dt$$
$$< \frac{\sqrt{z}/w_o}{k_o - 1 - \sqrt{z}/w_o} \int_{\sqrt{z}/w_o}^{k_o-1} t^{k_o-1} \exp(-t) dt$$
$$< \frac{\sqrt{z}/w_o}{k_o - 1 - \sqrt{z}/w_o} \Gamma(k_o),$$
(16)

where all the inequalities hold due to the monotonic increase of the function $f(z) = \left(\sqrt{z}/w_o\right)^{k_o-1} \exp\left(-\sqrt{z}/w_o\right)$ as $k_o$ is sufficiently large. We rewrite (16) to an equivalent form as

$$\frac{1}{\Gamma(k_o)} \gamma\left(k_o, \sqrt{z}/w_o\right) < \frac{\sqrt{z}/w_o}{k_o - 1 - \sqrt{z}/w_o}, \quad (17)$$

and the approximation is obtained as in the corollary by utilizing the asymptotic property in (11). □

For a given SNR, Corollary 1 demonstrates that the coverage probability scales up with $N^2$ and $N$ for the case of optimal and arbitrary phase shifts, respectively. Furthermore, the coverage probabilities converge to one as the number of phase shifts goes without bound.

### B. Ergodic Capacity

We now investigate the ergodic capacity, which is independent of small-scale fading on a long timescale. From Lemma 1, we first formulate the ergodic capacity with either arbitrary or the optimal phase shifts from the definition as

$$\bar{R}_a = \mathbb{E}\left\{\log_2\left(1 + \frac{\rho}{\sigma^2}\left|h_{\text{sd}} + \mathbf{h}_{\text{sr}}^H \boldsymbol{\Theta} \mathbf{h}_{\text{rd}}\right|^2\right)\right\}, \quad (18)$$

$$\bar{R}_o = \mathbb{E}\left\{\log_2\left(1 + \frac{\rho}{\sigma^2}\left(|h_{\text{sd}}| + \sum_{n=1}^N |[\mathbf{h}_{\text{sr}}]_n||[\mathbf{h}_{\text{rd}}]_n|\right)^2\right)\right\}, \quad (19)$$

One naive approach to evaluate the ergodic capacities in (18) and (19) is averaging over many different instantaneous channel realizations, but it requires high computational complexity and especially being burdensome with a large IRS array. Overcoming this issue, we compute the ergodic capacity in closed form.

**Theorem 2.** *For arbitrary phase shifts, the ergodic capacity [b/s/Hz] is*

$$\bar{R}_a = \frac{1}{\Gamma(k_a) \ln(2)} G_{2,3}^{3,1}\left(\frac{\sigma^2}{\rho w_a} \bigg| \begin{matrix} 0, 1 \\ 0, 0, k_a \end{matrix}\right), \quad (20)$$

*where $G_{p,q}^{m,n}\left(z \bigg| \begin{matrix} a_1, \ldots, a_p \\ b_1, \ldots, b_q \end{matrix}\right)$ is the MeijerG function [11, (9.301)]. Besides, for the optimal phase shifts, the ergodic capacity [b/s/Hz] is*

$$\bar{R}_o = \frac{2^{k_o - 1/2}}{\Gamma(k_o) \ln(2)} \frac{1}{\sqrt{2\pi}} G_{3,5}^{5,1}\left(\frac{\sigma^2}{4\rho w_o^2} \bigg| \begin{matrix} 0, \frac{1}{2}, 1 \\ 0, 0, \frac{1}{2}, \frac{k_o}{2}, \frac{k_o+1}{2} \end{matrix}\right). \quad (21)$$

*Proof.* For arbitrary phase shifts, the ergodic capacity in (18) is expressed as

$$\bar{R}_a = \int_0^\infty \log_2(1 + az) f_X(z) dz = \frac{1}{\ln 2} \int_0^\infty \ln(1 + az) f_X(z) dz, \quad (22)$$

where $X$ is defined in Appendix A and $f_X(z)$ is its probability density function (PDF). By a change of variable $t = az$, $a = \rho/\sigma^2$ and applying the integration by parts, (22) is equivalent to as

$$\bar{R}_a = \frac{1}{\ln 2} \int_0^\infty \frac{1}{1+t} (1 - F_X(t/a)) dt$$
$$= \frac{1}{\Gamma(k_a) \ln 2} \int_0^\infty \frac{1}{1+t} G_{1,2}^{2,0}\left(\frac{t\sigma^2}{\rho w_a} \bigg| \begin{matrix} 1 \\ 0, k_a \end{matrix}\right) dt, \quad (23)$$

where $F_X(\cdot)$ is the cumulative distribution function (CDF) of variable $X$. The last equality of (23) is obtained by an equivalence between the incomplete Gamma function and MeijerG function, i.e., $\Gamma(k_a, \gamma z/w_a) = G_{1,2}^{2,0}\left(\gamma z/w_a \bigg| \begin{matrix} 1 \\ 0, k_a \end{matrix}\right)$, with $\gamma = \sigma^2/\rho$ in our framework. After that, we obtain the result as shown in the theorem by computing the last integral in (23). Utilizing the above main steps, the ergodic capacity in (19) is computed for the optimal phase-shifts as

$$\bar{R}_o = \frac{1}{\Gamma(k_o) \ln 2} \int_0^\infty \frac{1}{1+z} \Gamma\left(k_o, \frac{\sqrt{z}\sigma}{\sqrt{\rho} w_o}\right) dz$$
$$= \frac{1}{\Gamma(k_o) \ln 2} \int_0^\infty \frac{1}{1+z} G_{1,2}^{2,0}\left(\frac{\sqrt{z}\sigma}{\sqrt{\rho} w_o} \bigg| \begin{matrix} 1 \\ 0, k_o \end{matrix}\right) dz, \quad (24)$$

then computing the last integral in (24), the ergodic capacity in case of the optimal phase shifts is obtained in the theorem. □

The results in Theorem 2 scale down the matters and are effectively used over many coherence intervals. To compare with previous works, we observe that the ergodic capacity in (18) and (19) has the inherent concavity regarding the SNR level. An upper bound is therefore obtained by Jensen's inequality together with the seminal results (27) and (37) in the appendix as

$$\bar{R}_a \leq \log_2\left(1 + \frac{\rho}{\sigma^2} \mathbb{E}\left\{\left|h_{\text{sd}} + \mathbf{h}_{\text{sr}}^H \boldsymbol{\Theta} \mathbf{h}_{\text{rd}}\right|^2\right\}\right)$$
$$= \log_2\left(1 + \frac{\rho}{\sigma^2}(\beta_{\text{sd}} + N\beta_{\text{sr}}\beta_{\text{rd}})\right), \quad (25)$$

$$\bar{R}_o \leq \log_2\left(1 + \frac{\rho}{\sigma^2} \mathbb{E}\left\{\left(|h_{\text{sd}}| + \sum_{n=1}^N |[\mathbf{h}_{\text{sr}}]_n||[\mathbf{h}_{\text{rd}}]_n|\right)^2\right\}\right)$$
$$= \log_2\left(1 + \frac{\rho}{\sigma^2}\left(\beta_{\text{sd}} + Nkw^2 + N^2k^2w^2 + Nkw\sqrt{\beta_{\text{sd}}\pi}\right)\right). \quad (26)$$

The upper bound for arbitrary phase shifts is scaling up with $N$, while that is $N^2$ in case of the optimal phase shifts. Even though these bounds on the ergodic capacity are established for a limited number of elements at IRS, they align with the power scaling law in the literature [5], [12]. To the end, Remark 1 addresses an extension of the proposed framework in this paper to other propagation channels with different communication scenarios.

**Remark 1.** *All the closed-form expressions on the coverage probability and ergodic capacity together with its upper*

bound are obtained for uncorrelated Rayleigh fading channels, which are tractable and matched well with rich-scattering propagation environments [13]. The analysis based on the moment-matching method can be easily extended to other fading channels, for example spatially correlated fading [14], where the moments of random variables (up to the forth order) are bounded. Moreover, by exploiting a linear beamforming technique, the framework in this paper can be extended to multi-antenna and/or multi-user scenarios.

## IV. Numerical Results

All above analyses are numerically verified by using the setup in [5], where the locations are mapped into a Cartesian coordinate system. Particularly, source is located at $(0, 0)$ and IRS is at $(40, 10)$. The large-scale-fading coefficients [dB] are modeled, similar to [14], [15], by $\beta_\mu = G_t + G_r + 10\nu_\mu \log_{10}(d_\mu/1\text{m}) - 30 + z_\mu$, where $\mu \in \{\text{sd}, \text{sr}, \text{rd}\}$; $G_t = 3.2$ dBi and $G_r = 1.3$ dBi are the antenna gain at the transmitter and receiver, respectively; $d_\mu$ is the distance between transmitter and receiver, while $\nu_\mu$ is the path loss exponent, i.e., $\nu_{\text{sd}} = \nu_{\text{sr}} = \nu_{\text{rd}} = 3.76$; the shadow fading is $z_\mu$ with the concrete setting clarified subsequently. The system uses 10 MHz bandwidth. The noise variance is $-94$ dBm corresponds to the noise figure 10 dB. The transmit power of source is 10 dBm and for the arbitrary phase shifts, the phase of each phase-shift element follows a uniform distribution. For comparison, a system without the assistance of IRS is included as a benchmark.

Fig. 1 plots the coverage probability versus different required information rates, where destination's location is $(60, 0)$, while shadow fading is $z_\mu = 0$ and the number of phase shifts is $N = 800$. The optimal phase shifts provide significantly better the coverage probability than a random phase-shift selection, especially when $\xi < 4$ [b/s/Hz]. Besides, the correctness of the closed-form expressions obtained in Theorem 1 can be observed as they overlap with Monte-Carlo simulations for all tested values. In this setting, an IRS with random phase shifts does not give any support to the system and gives the same performance as a traditional SISO system since the randomness may add up waveforms either constructively or destructively.

Fig. 2 displays the channel capacity [b/s/Hz] for the scenario, where the $x$-coordinate of destination is uniformly located in the range $[50, 200]$ and the shadow fading follows a Gaussian distribution with zero mean and standard derivation 4 dB. The closed-form expressions in Theorem 2 coincide with numerical simulations. Moreover, utilizing an IRS with random phase shifts always gives the same channel capacity as a traditional SISO system regardless the number of phase shifts. Finally, the superior improvement is obtained by utilizing the optimal phase shifts instead of an arbitrary set (up to 2.7×). The improvement becomes widely when integrating more reflecting elements.

## V. Conclusion

This paper has used an incomplete Gamma distribution and statistical information of Rayleigh channels to derive the closed-form expression of the coverage probability for IRS-enhanced communication with either the optimal or arbitrary phase shifts. We have further derived the ergodic capacity in a closed form based on a MeijerG function and found an upper bound with simple arithmetic operators only. While all the obtained closed-form expressions coincide with Monte-Carlo simulations, the asymptotic analysis and the upper bound on channel capacity unveil the scaling law of the number of phase shifts at IRS.

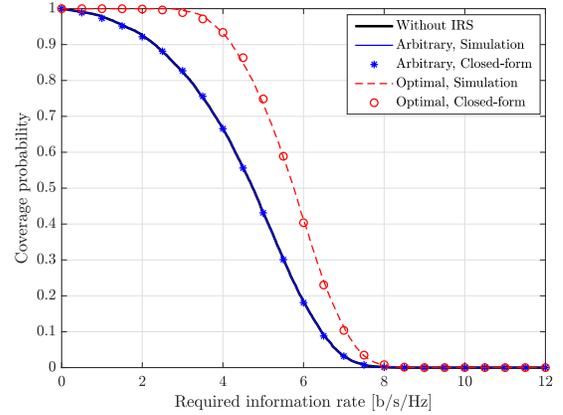

Fig. 1: The coverage probability versus the required information rate [b/s/Hz].

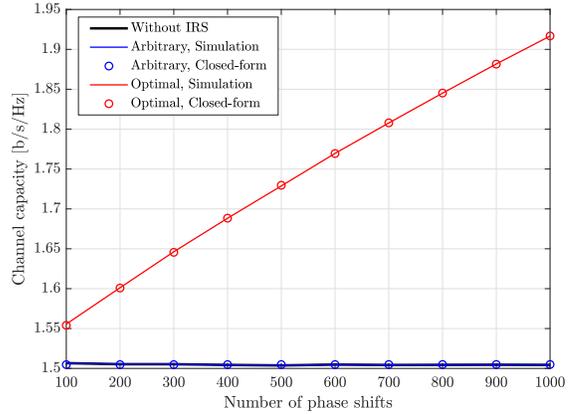

Fig. 2: The channel capacity [b/s/Hz] versus the number of phase shifts.

## Appendix

### A. Proof of Theorem 1 for Arbitrary Phase Shifts

Let us define $X = \left|h_{\text{sd}} + \mathbf{h}_{\text{sr}}^H \boldsymbol{\Theta} \mathbf{h}_{\text{rd}}\right|^2$, then its mean is obtained by the independence of channels as

$$\mathbb{E}\{X\} = \mathbb{E}\left\{|h_{\text{sd}}|^2\right\} + \mathbb{E}\left\{\left|\mathbf{h}_{\text{sr}}^H \boldsymbol{\Theta} \mathbf{h}_{\text{rd}}\right|^2\right\} = \beta_{\text{sd}} + N\beta_{\text{sr}}\beta_{\text{rd}}, \quad (27)$$

which is obtained due to the circular symmetry properties. We now compute $\mathbb{E}\{|X|^2\}$ as

$$\mathbb{E}\{|X|^2\} = \mathbb{E}\left\{\left|h_{\text{sd}} + \mathbf{h}_{\text{sr}}^H \boldsymbol{\Theta} \mathbf{h}_{\text{rd}}\right|^4\right\}$$
$$= \mathbb{E}\left\{\left|\underbrace{|h_{\text{sd}}|^2}_{a} + \underbrace{h_{\text{sd}}^* \mathbf{h}_{\text{sr}}^H \boldsymbol{\Theta} \mathbf{h}_{\text{rd}}}_{b} + \underbrace{h_{\text{sd}} \mathbf{h}_{\text{rd}}^H \boldsymbol{\Theta}^H \mathbf{h}_{\text{sr}}}_{c} + \underbrace{\left|\mathbf{h}_{\text{sr}}^H \boldsymbol{\Theta} \mathbf{h}_{\text{rd}}\right|^2}_{d}\right|^2\right\}$$
$$= \mathbb{E}\left\{|a|^2\right\} + \mathbb{E}\left\{|b|^2\right\} + \mathbb{E}\left\{|c|^2\right\} + 2\mathbb{E}\{ad\} + \mathbb{E}\left\{|d|^2\right\},$$
$$(28)$$

where $\mathbb{E}\left\{|a|^2\right\} = 2\beta_{\text{sd}}^2$ by virtue of [16, Lemma 2.9], $\mathbb{E}\left\{|b|^2\right\} = \mathbb{E}\left\{|c|^2\right\} = N\beta_{\text{sd}}\beta_{\text{sr}}\beta_{\text{rd}}$ and $\mathbb{E}\{ad\} = N\beta_{\text{sd}}\beta_{\text{sr}}\beta_{\text{rd}}$

by the independence of the channels, and the last expectation of (28) is computed as

$$\mathbb{E}\left\{|d|^2\right\} = \mathbb{E}\left\{\|\boldsymbol{\Theta}\mathbf{h}_{\text{rd}}\|^4 \left|\frac{\mathbf{h}_{\text{sr}}^H\boldsymbol{\Theta}\mathbf{h}_{\text{rd}}}{\|\boldsymbol{\Theta}\mathbf{h}_{\text{rd}}\|} \frac{\mathbf{h}_{\text{rd}}^H\boldsymbol{\Theta}^H\mathbf{h}_{\text{sr}}}{\|\boldsymbol{\Theta}\mathbf{h}_{\text{rd}}\|}\right|^2\right\}. \quad (29)$$

Let us define $z = \mathbf{h}_{\text{sr}}^H\boldsymbol{\Theta}\mathbf{h}_{\text{rd}}/\|\boldsymbol{\Theta}\mathbf{h}_{\text{rd}}\|$, then $z \sim \mathcal{CN}(0, \beta_{\text{sr}})$ and (29) is manipulated as

$$\mathbb{E}\left\{|d|^2\right\} = \mathbb{E}\left\{\|\boldsymbol{\Theta}\mathbf{h}_{\text{rd}}\|^4 |z|^4\right\} \stackrel{(a)}{=} \mathbb{E}\left\{\|\boldsymbol{\Theta}\mathbf{h}_{\text{rd}}\|^4\right\}\mathbb{E}\left\{|z|^4\right\}$$
$$\stackrel{(b)}{=} N(N+1)\beta_{\text{rd}}^2 2\beta_{\text{sr}}^2 = (2N^2 + 2N)\beta_{\text{rd}}^2\beta_{\text{sr}}^2, \quad (30)$$

where $(a)$ is obtained by the fact that $\boldsymbol{\Theta}\mathbf{h}_{\text{rd}}$ and $z$ are independent; $(b)$ is obtained by the facts that $\boldsymbol{\Theta}\mathbf{h}_{\text{rd}} \sim \mathcal{CN}(\mathbf{0}, \beta_{\text{rd}}\mathbf{I}_N)$ and the use of [17, Lemma 9] to compute the forth moment of symmetric complex Gaussian random variables. Combining the results of (28)-(30), we obtain

$$\mathbb{E}\{|X|^2\} = 2\beta_{\text{sd}}^2 + 4N\beta_{\text{sd}}\beta_{\text{sr}}\beta_{\text{rd}} + (2N^2 + 2N)\beta_{\text{rd}}^2\beta_{\text{sr}}^2. \quad (31)$$

By utilizing the identity $\text{Var}\{X\} = \mathbb{E}\{|X|^2\} - |\mathbb{E}\{X\}|^2$ together with the results in (27) and (31), we obtain

$$\text{Var}\{X\} = \beta_{\text{sd}}^2 + 2N\beta_{\text{sd}}\beta_{\text{sr}}\beta_{\text{rd}} + (N^2 + 2N)\beta_{\text{sr}}^2\beta_{\text{rd}}^2. \quad (32)$$

We now use the moment-matching method to match the CDF of $X$ to a Gamma distribution with the shape and scale parameters are computed as $k_a = \frac{(\mathbb{E}\{X\})^2}{\text{Var}\{X\}}, w_a = \frac{\text{Var}\{X\}}{\mathbb{E}\{X\}}$. Then by exploiting (27) and (32) into these equations, we obtain the result as shown in the theorem.

### B. Proof of Theorem 1 for the Optimal Phase Shifts

From the optimal channel capacity in (3), we first define $A = |h_{\text{sd}}|$ and $B_n = |[\mathbf{h}_{\text{sr}}]_n||[\mathbf{h}_{\text{rd}}]_n|, \forall n$. Variable $A$ follows Rayleigh distribution with mean $\sqrt{\beta_{\text{sd}}\pi}/2$ and variance $(4-\pi)\beta_{\text{sd}}/4$, thus $\mathbb{E}\{A^2\} = \text{Var}\{A\} + (\mathbb{E}\{A\})^2 = \beta_{\text{sd}}$. Notice that $|[\mathbf{h}_{\text{sr}}]_n|$ and $|[\mathbf{h}_{\text{rd}}]_n|$ also follow Rayleigh distribution with CDF and PDF as

$$F_{[\mathbf{h}_\alpha]_n}(z) = 1 - \exp\left(-\frac{z^2}{\beta_\alpha}\right), f_{[\mathbf{h}_\alpha]_n}(z) = \frac{2z}{\beta_\alpha}\exp\left(-\frac{z^2}{\beta_\alpha}\right), \quad (33)$$

where $\alpha \in \{\text{sr}, \text{rd}\}$. The CDF of $B_n$, denoted by $F_{B_n}(z)$, is

$$F_{B_n}(z) = \Pr(B_n < z) = \Pr\left(|[\mathbf{h}_{\text{sr}}]_n| < \frac{z}{|[\mathbf{h}_{\text{rd}}]_n|} = \frac{z}{v}\right)$$
$$= \int_0^\infty F_{[\mathbf{h}_{\text{sr}}]_n}(z/v) f_{[\mathbf{h}_{\text{rd}}]_n = v}(v) dv = 1 - \frac{2z}{\sqrt{\beta_{\text{sr}}\beta_{\text{rd}}}} K_1\left(\frac{2z}{\sqrt{\beta_{\text{sr}}\beta_{\text{rd}}}}\right), \quad (34)$$

where $K_n(\cdot)$ is the modified Bessel function of the second kind with $n$-th order. The PDF of $B_n$ is then computed as

$$f_{B_n}(z) = \frac{dF_{B_n}(z)}{dz} = \frac{2z}{\beta_{\text{sr}}\beta_{\text{rd}}}\left(K_2\left(\frac{2z}{\sqrt{\beta_{\text{sr}}\beta_{\text{rd}}}}\right) + K_0\left(\frac{2z}{\sqrt{\beta_{\text{sr}}\beta_{\text{rd}}}}\right)\right)$$
$$- \frac{2}{\sqrt{\beta_{\text{sr}}\beta_{\text{rd}}}}K_1\left(\frac{2z}{\sqrt{\beta_{\text{sr}}\beta_{\text{rd}}}}\right), \quad (35)$$

thanks to [11, (8.486.11)]. By utilizing the PDF expression in (35), we can compute the statistical information of $B_n$ as

$$\mathbb{E}\{B_n\} = \pi\sqrt{\beta_{\text{sr}}\beta_{\text{rd}}}/4, \text{Var}\{B_n\} = \left(1 - \pi^2/16\right)\beta_{\text{sr}}\beta_{\text{rd}}. \quad (36)$$

Let us deploy the moment-matching method such that $B_n$ is matched to a Gamma distribution, for which the shape parameter $k$ and the scale parameter $w$ satisfy $kw = \mathbb{E}\{B_n\}$ and $kw^2 = \text{Var}\{B_n\}$. Subsequently, we obtain those parameters as shown in the theorem. We now define $C = A + \sum_{n=1}^N B_n$, which has mean and second moment are computed as

$$\mathbb{E}\{C\} = \frac{\sqrt{\pi\beta_{\text{sd}}}}{2} + Nkw, \mathbb{E}\{C^2\} = \beta_{\text{sd}} + Nkw(w + Nkw + \sqrt{\beta_{\text{sd}}\pi}), \quad (37)$$

and its variance is computed as

$$\text{Var}\{C\} = \mathbb{E}\{C^2\} - (\mathbb{E}\{C\})^2 = \beta_{\text{sd}} + Nkw^2 - \beta_{\text{sd}}\pi/4. \quad (38)$$

Let us match random variable $C$ to a Gamma distribution whose the shape and scale parameters are defined as in (9). By introducing a new random variable $D = C^2$, then the CDF of $D$ can be directly drawn from $C$ as $F_D(x) = F_C(\sqrt{x})$ and we close the proof here.